\documentclass[useAMS,usenatbib]{mn2e}
\usepackage{natbib}
\usepackage{graphicx}

\title[AGB effect on $\omega$~Cen]{Test on the AGB Scenario as the Origin
of the Extreme-Helium Population in $\omega$~Centauri}
\author[E. Choi \& S. K. Yi]{Ena Choi and Sukyoung K. Yi \\
Department of Astronomy,
Centre for Space Astrophysics, Yonsei University, Seoul 120-749, Korea}

\begin{document}

\pagerange{\pageref{firstpage}--\pageref{lastpage}} \pubyear{2007}

\maketitle

\label{firstpage}


\begin{abstract}

The most massive Galactic globular cluster, $\omega$~Centauri, appears
to have multiple populations. Its bluest main sequence and
extreme horizontal branch stars are suggested to have the common
origin, that is, an extremely high helium abundance of $Y \sim 0.4$.
The high helium abundance is most often attributed to asymptotic
giant branch (AGB) stars. In this study we test the AGB hypothesis.
We simulate the maximum-AGB models where the impact of AGB stars
is maximised by assuming that supernova explosions do not affect the
chemical evolution of the proto cloud.
We compare the enrichment history of helium, metals, carbon
and nitrogen to the observed values. Even under the most generous condition,
the maximum-AGB models fail to reproduce the large values of helium
$Y \sim 0.4$ and helium enrichment parameter
$\Delta Y / \Delta Z \sim 70$ which were deduced
from the colour-magnitude diagram fits.
They also fail to reproduce the C and N contents of the blue population
spectroscopically determined.
We conclude that the AGB scenario with the canonical stellar
evolution theory cannot explain the observational constraints and that
the self chemical enrichment does not provide a viable solution.
Alternative processes are desperately called for.
\end{abstract}



\section{Introduction}

The origin of the blue population in the most massive Galactic
globular cluster (GC), $\omega$~Centauri, is at
the centre of debate after the discovery of its multiple
populations \citep[e.g.][]{a97,lee99,b04}.
The explanation of the blue main sequence (bMS) of $\omega$~Cen
by a huge excess of the primordial helium abundance, $\Delta$$Y\sim 1.2$,
was implied via the colour-magnitude diagram (CMD) fitting \citep{n04}.
This bMS, which contains $\sim$ 30 percent of the cluster stars,
is roughly 0.3 dex more metal-rich than the red main sequence (rMS)
population \citep{pi05}.
Its blue colour despite its high metallicity is attributed to an
extremely high value of helium.
This explanation has further been reinforced
by the fact that the same helium excess can reproduce the extended
horizontal branch (EHB) of $\omega$~Cen \citep{dc04,lee05}.
Follow-up spectroscopic and photometric investigations have been performed
by many groups \citep[e.g.][]{kay05, sol06, sta07,vil07,van07}.
The extreme helium scenario appears to provide a successful solution
to another cluster, NGC\,2808, which also shows multiple main sequences
\citep{pi07} and a pronounced EHB \citep{lee05}.
It has recently been claimed that the GCs with an EHB are among
the most massive in the Milky Way, and the possible link between the
mass of the cluster and the helium anomaly is heavily investigated on
\citep{rb06,lee07}.
The existence of helium-rich GCs in the external galaxy, M\,87, has
also been suggested \citep{kav07}, implying that this phenomenon
may be universal.

The extreme helium scenario, however, has been criticised.
This is mainly because the helium-enrichment parameter $\Delta Y / \Delta Z $
corresponding to proposed helium abundance $\sim$~70 or even higher,
which is more than an order of magnitude larger than observational and
theoretical values, $\Delta Y / \Delta Z =1$ -- 5
\citep[e.g.][]{fer96,pp98,ji03}.
For possible origins for the extreme helium abundance, asymptotic 
giant branch (AGB) stars, massive stars, and Type II supernovae (SN II)
have been widely discussed \citep[]{n04,dan05,mm06}.
\citet{bn06} however demonstrated that such candidates cannot produce
the amount of helium required, for ordinary initial mass functions (IMFs)
within the scheme of a closed-box self enrichment.
More recently, \citet{cy07} showed that essentially no population
can produce such a high value of $\Delta Y / \Delta Z $ via self-enrichment
processes regardless of the shape of IMF.

Supernovae are not helpful for generating a high value of
$\Delta Y / \Delta Z $ because they produce metals as well as helium.
The ejecta from AGB stars is generally believed to show the largest
helium enrichment parameters \citep[e.g., see][]{Izz04, cy07}.
However, the AGB hypothesis is found to be unsuccessful for matching
both the high helium abundance and the CNO properties constrained from
observations even when the supernova effect is removed from
the calculation \citep{ka06,b07}.

We further elaborate on the works of \citet{ka06,b07} by testing
{\em the maximum-AGB hypothesis} where the impact of AGB stars is maximised
by assuming that massive star ejecta of unknown mass range 
escape the small potential wells of proto clouds.
In this letter we search for the parameter space that satisfies all
the observational constraints including spectroscopic measurements.

\section{The chemical Evolution Models}
\subsection{Introduction}
We adopt the chemical enrichment code described in \citet{cy07}
following the formalism of \citet{tin80}. The
two-component model assuming instant mixing and cooling
traces the net metallicity $Z$, helium $Y$, carbon and nitrogen
contents of the gas and stellar components.

The stellar mass, $M_s(t)$, and the cold gas mass, $M_g(t)$, are
normalised to the initial system mass,
\begin{equation}
\mu_s(t)\equiv {M_s(t)\over M_{tot}(0)},~~~\mu_g(t)\equiv
{M_{g}(t)\over M_{tot}(0)}.
\end{equation}

In this study, we assume that the formation of the former (rMS)
population is essentially instant:
\begin{equation}
\psi(t \neq 0)=0.
\end{equation}
The evolution of the gas mass is given by
\begin{equation}
{d\mu_g \over dt} = E(t)-\psi(t)
\end{equation}
where the ejecta gas mass at time $t$, $E(t)$,  is defined as
\begin{equation}
E(t) = \int_{m_t}^{m_{upper}} dm\phi (m)(m-w_m)\psi(t-\tau_m).
\end{equation}
where $\phi(m)$ denotes the initial mass function (IMF), $w_m$ the remnant
mass for a star with main sequence mass $m$, $\tau_m$ the lifetime of 
a star of mass $m$, and $m_{upper}$ the upper mass cut
in the initial mass function. The Scalo IMF \citep{s86} with cutoffs at 0.1
and 100$M_{\odot}$ is assumed, and the remnant mass $w_m$ and the lifetime
of a star $\tau_m$ are adopted from \citet{fs00, fs01}.

The equation for the evolution of metallicity in gas is given by
\begin{equation}
{d(Z_g\mu_g) \over dt} =-\psi(t)+E_Z(t)
\end{equation}
where the mass of ejected metal at time $t$, $E_Z(t)$, is defined as
\begin{eqnarray}
E_Z(t) =\int_{m_t}^{m_{upper}} dm\phi (m) \big[mp_m\psi(t-\tau_m)
 +(m-w_m) \psi (t-\tau_m) Z_g (t-\tau_m)\big]
\end{eqnarray}
where $p_m$ denotes the newly synthesized and ejected metal (or helium, C,
N) mass fraction to initial stellar mass. We will discuss this in greater
detail in \S 2.2.

The initial chemical properties of the models are adopted from the
observational values for the rMS population of $\omega$~Cen.
The initial metallicity $Z_0 = 0.001$ and the initial C, O abundance
[C/M]$=0.0$, [N/M]$=1.0$ are from the spectroscopic measurements of
\citet{pi05}, and the initial helium abundance $Y_0 = 0.232 $ is deduced
from the CMD fit in \citet{lee05}.

\subsection{Chemical Yields}
The chemical yields ($p_m$), defined as the mass fraction of a
star of mass $m$ that is $newly$ converted to metals or helium
and ejected, are the most important input parameter to the chemical
evolution of a population. We basically adopt the helium, carbon and metal
$p_m$ predictions of \citet{m92} for the mass range $9 \-- 100 M_{\odot}$.
For nitrogen, we adopt the $p_m$ prediction of \citet{no97}.
For $M_{\odot} \leq 6$ we adopt the helium, carbon, nitrogen, and metal
yields from \citet{her04}.  We confirm that the use of different yields
\citep[e.g.][]{van97, ven02} does not make any notable difference 
to our conclusion. We use the chemical yields from the hermitian fits
to the $p_m$ predictions. In Figure 1, we show the
theoretical ejecta abundance predicted by our fitting functions from
the metal-poor ($Z = 0.001$) stars of mass $0.1 \-- 100 M_{\odot}$.
The chemical abundances predicted and the resulting
helium enrichment parameters are presented. As shown in the middle panel,
the highest peak around $6\,M_{\odot}$ in the helium enrichment parameter diagram
reaches the value proposed for the bMS population.
Indeed, the AGB ejecta appears to be the most promising candidate.
Supernovae on the other hand produce more metals than helium and hence
are not helpful for generating a high helium-to-metal population.

\begin{figure}
\begin{center}
\includegraphics[width=\columnwidth]{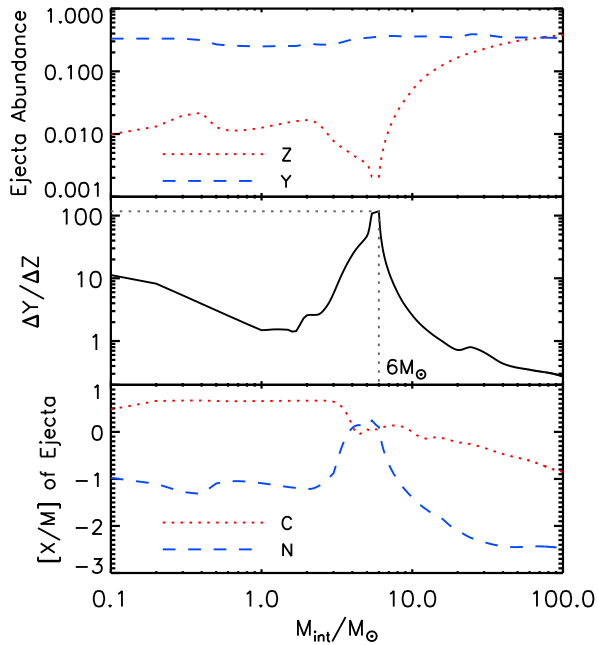}
\caption{ The theoretical ejecta abundance calculated from the stellar yields
of \citet{m92} and \citet{her04}. {\em Top:} The helium and
metal abundances of the ejecta from the metal-poor ($Z = 0.001$) stars
of mass $0.1 \-- 100 M_{\odot}$. {\em Middle:} The resulting helium
enrichment parameters.
{\em Bottom:} The carbon and Nitrogen abundances of the ejecta from the
stars with the initial compositions [C/M] $=0.0$, [N/M] $=1.0$ following
\citet{pi05}.
 \label{fig1}}
\end{center}
\end{figure}

\subsection{The Maximum-AGB model}

It has already been shown that a closed-box system cannot achieve
such high values of helium and helium enrichment parameter as implied for the blue
population of $\omega$~Cen \citep{cy07}.
This is mainly because of the ill effect of supernovae on
$\Delta Y / \Delta Z $.

In realistic models for small potential wells, the remnants of supernovae
are believed to escape the system \citep{l74}.
Such explosions would probably remove the remnant gas in the potential
as well after the first star formation episode.
But in general, the escape of the supernova materials is likely to be
easier than the escape of the remnant gas.
So we set up {\em ad hoc}
models in which the supernova ejecta escape the system
without affecting the chemical properties of the remnant gas, hence
maximising the chemical influence of AGB stars.
We call this ``the maximum-AGB models''.

Our prescription has two input parameters.
The first free parameter, $m_{esc}$, indicates the critical
mass above which the stellar mass ejecta escape the potential well
without affecting the chemical abundance of the remaining gas.
This parameter is likely a function of the mass of the system
because the potential well determines whether the supernova-driven
winds can escape the gravitational potential of the system or not.
\citet{ka06} assumed that the system can retain the ejecta
from the stars with $m \leq 6.5 M_{\odot}$. We explore an expanded
parameter space for $5 M_{\odot} \leq m_{esc} \leq 99 M_{\odot}$.
The scenario of Karakas et al. is equivalent to our
scenario with $m_{esc}=6.5 M_{\odot}$.
The model with the largest $m_{esc}$ represents the ordinary
chemical evolution model which takes account of all the mass ejecta
from stars when calculating the metallicity of the next generations.
The model with the smallest $m_{esc}$ shows the extreme condition that
the ejecta from AGB stars dominates the chemical properties of
the next generations, hence the maximum-AGB model.
Then we define $E(t)$, the ejecta gas mass at time $t$ as
\begin{equation}
E(t) = \int_{m_t}^{m_{esc}} dm\phi (m)(m-w_m)\psi(t-\tau_m),
\end{equation}
and $E_Z(t)$, the mass of the ejected metal at time $t$ as
\begin{equation}
\begin{array}{ll}
E_Z(t) =\int_{m_t}^{m_{esc}} dm\phi (m) \big[mp_m\psi(t-\tau_m)
 +(m-w_m)
        \psi(t-\tau_m) Z_g (t-\tau_m)\big].
\end{array}
\end{equation}

\begin{figure}
\begin{center}
\includegraphics[width=\columnwidth]{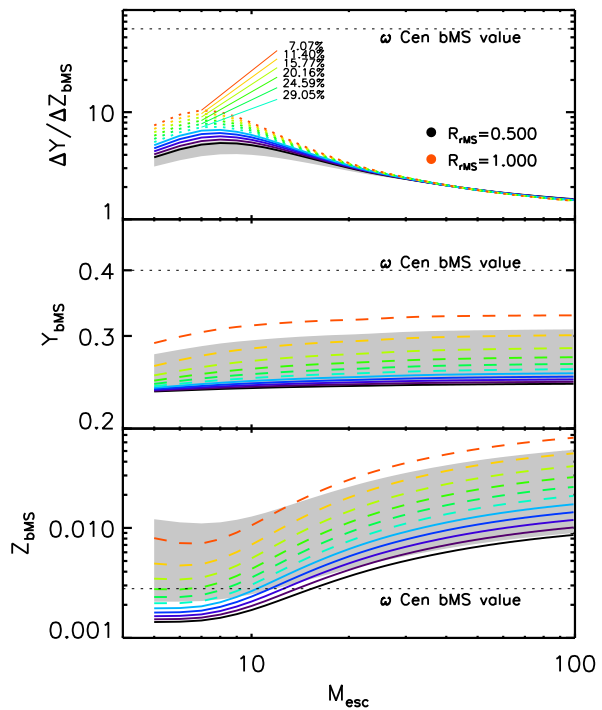}
\caption{The  chemical abundances of the second stellar generation resulting
from the maximum-AGB chemical evolution model. The results are shown
for different values of initial starburst mass fractions, $R_{rMS}=0.5$--1.0
with $\Delta$$R_{rMS}=0.05$. Dashed/solid lines show the models that are
incompatible/compatible with the empirical mass fraction of the blue population
(30 percent), for the age difference of 1\,Gyr. The shaded region shows the results
for the age difference of 3\,Gyr.
The chemical abundances of the blue population of $\omega$ Cen are
shown by the black dotted
horizontal line in each panel. {\em Top:} The helium enrichment parameter.
The mass fractions of the blue population are shown for the models that
are incompatible with the empirical requirement (30 percent). {\em Middle:}
The helium content. {\em Bottom:} The metallicity.
\label{fig2}}
\end{center}
\end{figure}

Another free parameter is $R_{rMS}$ which indicates the mass ratio
of the initial starburst i.e., for the former, red population of 
$\omega$~Cen to the total system mass.
We assume that the formation of the former (red) population is essentially
instant and regulate the fraction of the initial star formation by using the
free parameter $R_{rMS}$. We calculate models for $0.5 \leq
R_{rMS} \leq 1.0$. The model with $R_{rMS}= 1.0$ means all gas goes into the star
formation at time $t=0$ and the metallicity of the gas that forms the next
generation of stars is solely influenced by the ejecta from the former
generation. Then we define the star formation rate as
\begin{equation}
\psi(t) = R_{rMS} \ \delta(t)
\end{equation}
where $\delta(t)$ is a Dirac delta function.


In effect we are testing a self-enrichment hypothesis where
the helium enrichment of the blue population is solely due to the
ejecta from the former (red) population.
We calculate the
gas enrichment history for the predicted age difference
between the two extreme populations of $\omega$~Cen, i.e. $1 \-- 3$ Gyr
\citep{lee05, sta06}. We also consider the observed number fraction of the bMS
stars ($\sim 30 percent$) as an additional constraint.
The rMS-to-bMS ratio, 7:3, may not be a strong constraint if
the bMS and rMS populations formed in geographically different regions.
For example, the bMS stars are observed to be more centrally concentrated
than the rMS stars \citep{sol07}, and thus if one assumes that the
original rMS stars had a better chance of escaping from the system
potential during the dynamical encounters with the Milky Way galaxy
over the Hubble time, the original mass ratio may have been higher.
We discuss the impact of this uncertainty in \S 3.

\section{Results}

\begin{figure}
\begin{center}
\includegraphics[width=\columnwidth]{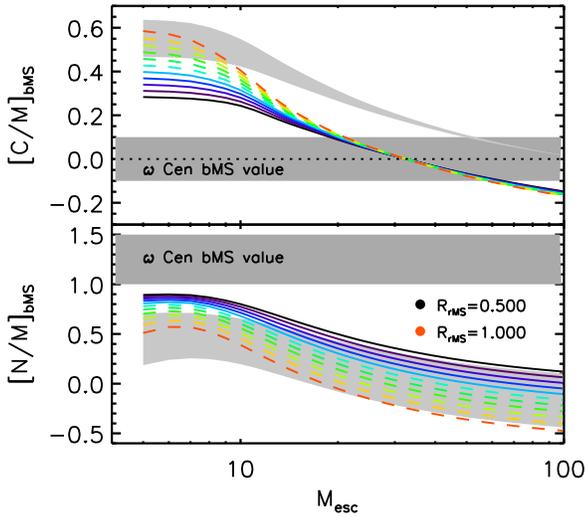}
\caption{ The same as Figure 2, but for the spectroscopic measurements
[C/M] {\em (top)} and [N/M] {\em (bottom)}. \label{fig3}}
\end{center}
\end{figure}

Figure 2 shows the helium and metal contents and the helium enrichment
parameter of
the remaining gas after a 1-Gyr evolution.
The chemical properties of the models are presented for our
$m_{esc}$ -- $R_{rMS}$ parameter space.
The models assuming $\Delta$$t \sim 1 $ Gyr (that is, the age difference
between the old red and the young blue populations) are shown by lines,
while the shaded region shows the results for $\Delta$$t \sim 3 $ Gyr which
is close to the upper limit on the age difference between the two populations
\citep{sta06}.
The solid/dashed lines show the models that are compatible/incompatible
with the observed mass ratio, 7:3, respectively.
As shown in the top panel of Fig 2, the model with
$m_{esc}=7 M_{\odot}$ and $R_{rMS}=1.0$ has the largest value of
$\Delta Y / \Delta Z \sim 10.5$. This value is much higher than the
canonical value, $\Delta Y / \Delta Z \sim 2$ \citep{pp98}
yet still much lower than the value we are seeking for
(i.e., $\Delta Y / \Delta Z \sim 70$). In addition, the maximum helium
content which can be achieved in this model is Y $\sim 0.33$ (for
$m_{esc} \ge 50 M_{\odot}$ and $R_{rMS}=1.0$), also much smaller than Y $\sim
0.4$ estimated by the CMD analysis of $\omega$~Cen. Even when we assume
an extreme condition, $m_{esc}=5\,M_{\odot}$, that is hardly plausible in
terms of physics, the extremely high values of helium and
helium enrichment parameter suggested for the bMS cannot be reproduced.

\begin{figure}
\begin{center}
\includegraphics[width=\columnwidth]{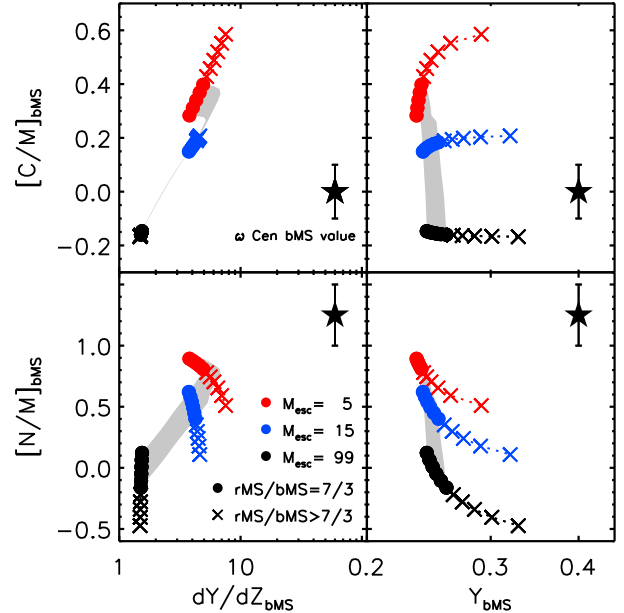}
\caption{ The calculated chemical abundance of the ejecta for the
entire ($R_{rMS}=0.5 \sim 1.0$, $\Delta R = 0.05$) parameter spaces
of the maximum-AGB model is represented.  Filled
circle and cross represents the resulting value of $m_{esc}=5, 15, 99 M_{\odot}$
model. Filled circle indicates that the condition of mass ratio
between rMS and bMS is satisfied while cross denotes the model which
do not meet this condition. Shaded region shows the viable region of
the total parameter space, $m_{esc}=5 \-- 99 M_{\odot}$, which satisfies the
mass ratio condition. Reference values of blue population of
$\omega$~Cen are marked(filled star) in each panel. \label{fig4}}
\end{center}
\end{figure}

As mentioned in \S 2, the mass fraction of the bMS population
(30 percent) may be an upper limit. As seen in the top and middle panels of
Figure 2, models cannot reach the required values even when we lower the
bMS mass fraction constraint by a significant factor.
The situation becomes even worse
when we compare the models to the spectroscopic line strengths.
In Figure 3, we present the resulting carbon and nitrogen contents of
the models for the same $m_{esc}$ and $R_{rMS}$ parameter space as in
Figure 2. This result is important since the reference chemical
properties of the bMS are directly measured {\em via spectroscopy},
while the helium abundance and the helium enrichment parameter
are {\em deduced} from CMD fits.
The models with $m_{esc} \approx 7 M_{\odot}$ show the largest value of
$\Delta Y / \Delta Z $ (Figure 2 top panel) but do not match the
observed carbon content of the bMS (Figure 3 top panel).
The horizontal, shaded band shows the observed value for the bMS stars
\citep{pi05}. However, the errors are not given and can be
more significant than we estimated from the literature \citep{pi05, sta07}.
In case of the red population, [N/M] is poorly constrained;
\citet{pi05} mentioned that values of [N/M] $\leq 1.0$ could also be compatible.
We set [N/M] $=1$ for the red population.
This assumption on the nitrogen abundance of the rMS population
may appear unphysically high considering the typical nitrogen production
of stars, that is, an order of [N/M]$\sim -1$ through 0, as shown in Figure 1.
The use of lower values of [N/M]$_{rMS}$ results in a substantially greater
mismatch in [N/M]$_{bMS}$ in this figure.

We present the resulting carbon, nitrogen and helium abundances
simultaneously in Figure 4. The three cases of $m_{esc}= 5, 15, 99 M_{\odot}$
are shown. Filled circle (cross) show the models that are consistent
(inconsistent) with the mass ratio constraint, respectively.
The shaded region shows the the parameter space that satisfies
the mass ratio constraint.  The observed chemical properties of the
blue population stars of $\omega$~Cen (filled star in each panel)
are not reproduced under any circumstance. Note the logarithmic scale of
the axes.

\begin{figure}
\begin{center}
\includegraphics[width=\columnwidth]{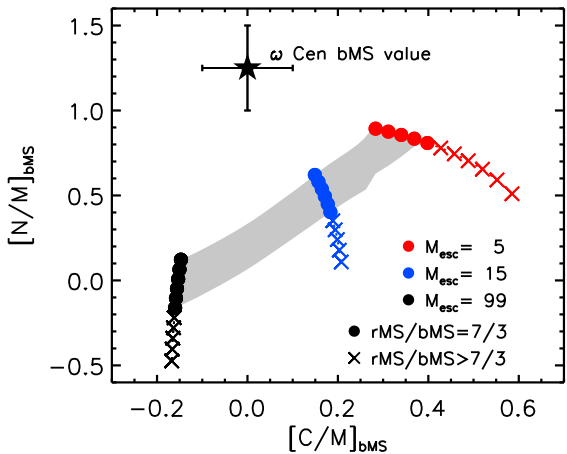}
\caption{ The same as Figure 4 but for the [C/M] and [N/M] space.\label{fig5}}
\end{center}
\end{figure}

Figure 5 shows that the models fail to match the {\em spectroscopic}
data of \citet{pi05}. The empirical value (star symbol with errors) 
measured using the VLT is in reasonable agreement with the more recent
measurements of \citet{sta07}. As mentioned, the models are based on
the assumption of [C/M] = 0.0 and [N/M] = 1.0 for the red population. 
The model values of [N/M]$_{bMS}$ may be considered upper limits.
The mismatch is significant.

\section{Discussion}

We have used a chemical enrichment model to demonstrate that the
AGB ejecta of the red population cannot have produced the blue
population of $\omega$~Cen.
We tested a hypothesis in which the massive star ejecta escape the potential
well so that the AGB effect can be maximised.
But even in this {\em ad hoc} scenario, models miserably fail to
match the chemical properties of the blue population of $\omega$~Cen.
We use the bMS mass fraction (30 percent) and the age difference
between the two populations (1--3\,Gyr) as constraints, but changing them
does not make the match any better.
We as a result confirm the result of \citet{ka06}.

The helium content of the AGB ejecta is basically too small to be
consistent with the value suggested for the blue population of
$\omega$~Cen, Y$\sim 0.4$, as we discussed earlier \citep{cy07}.
\citet{Ro07} has also confirmed this using various yields from
independent studies \citep[e.g.][]{van97,mm02,hi06,mey06}.
Such a high level of helium excess can be achieved only through
massive stars \citep{n04,dan05,mm06}. However, such massive stars
also produce a large amount of metals and thus resulting in
a small value of  $\Delta Y / \Delta Z $ \citep{cy07}.
In this respect, a variation in the initial mass function is useless.
The escape of the massive star ejecta through supernova-driven
winds would lead to a larger $\Delta Y / \Delta Z $ roughly up to 10;
yet, still much lower than suggested by the CMD fits.
Even if we admit that the estimation for the helium abundance via CMD fits
is extremely difficult and so uncertain, the maximum-AGB scenario cannot 
reproduce the carbon and nitrogen properties spectroscopically observed
and hence probably more robust, either.

For an alternative solution, surface pollution scenarios \citep[]{Tsu07,nt07}
have been suggested and deserve attention especially because
they demand only mildly-enhanced helium abundance for the blue population.
From a different avenue, \citet{chu06} suggested that the primordial
helium may have been concentrated in stellar-mass scales within
minihalos of size roughly consistent with dwarf galaxies.
This aspect is particularly appealing because $\omega$~Cen is often
considered as a remnant of a former dwarf galaxy and the multiple
main sequences are found in the most massive globular clusters in general.
This scenario offers an attractive solution to the helium variation in
$first$ stars. However, it remains unclear how long the
extreme-helium clumps could last. The blue population of $\omega$~Cen
appears to be substantially enhanced in metals as well and hence
hardly qualifies as first stars and younger than the red population 
by a couple of billion years. 
\citet{cy07} proposed the fluctuation in the chemical properties
of the initial starburst clumps during and after the first star formation
epoch. Given the impact of the issue to the understanding of the
galaxy formation, a more detailed and realistic model calculation
is necessary until such scenarios become convincing.

\section*{Acknowledgments}
We are grateful to Ignacio Ferreras, Leonis Chuzhoy, Young-Wook Lee, 
Hansung Gim for useful comments.


\end{document}